\documentclass[conference]{IEEEtran}
\IEEEoverridecommandlockouts
\usepackage{cite}
\usepackage{amsmath,amssymb,amsfonts}
\usepackage{bbding}

\usepackage[colorlinks,allcolors=black,pdftex]{hyperref}
\usepackage{algorithmic}
\usepackage{graphicx}
\usepackage{textcomp}
\usepackage{xcolor}
\def\BibTeX{{\rm B\kern-.05em{\sc i\kern-.025em b}\kern-.08em
    T\kern-.1667em\lower.7ex\hbox{E}\kern-.125emX}}
\begin{document}

\title{Community Detection in Electrical Grids \\ Using Quantum Annealing
}

\author{\IEEEauthorblockN{Marina Fernández-Campoamor\IEEEauthorrefmark{1} \IEEEauthorrefmark{2},
Corey O'Meara\IEEEauthorrefmark{1},
Giorgio Cortiana\IEEEauthorrefmark{1}, 
Vedran Peric\IEEEauthorrefmark{2} and
Juan Bernabé-Moreno\IEEEauthorrefmark{1}
}
\IEEEauthorblockA{\IEEEauthorrefmark{1}E.ON Digital Technology GmbH\\
Munich, Germany
\\ Email: marina.fernandez-fernandez-campoamor@eon.com}
\IEEEauthorblockA{\IEEEauthorrefmark{2}Technical University of Munich\\
Munich, Germany}
}

\maketitle

\begin{abstract}
With the increase of intermittent renewable generation resources feeding into the electrical grid,  Distribution System Operators (DSOs) must find ways to incorporate these new actors and adapt the grid to ensure stability and enable flexibility. 
Dividing the grid into logical clusters entails several organization and technical benefits, helping overcome these challenges.
However, finding the optimal grid partitioning remains a challenging task due to its complexity. 
At the same time, a new technology has gained traction in the last decades for its promising speed-up potential in solving non-trivial combinatorial optimization problems: quantum computing. This work explores its application in Graph Partitioning using electrical modularity. We benchmarked several quantum annealing and hybrid methods on IEEE well-known test cases. The results obtained for the IEEE 14-bus test case show that quantum annealing \textit{DWaveSampler} brings equal solutions or, for the optimal number partitions, a 1\% improvement. For the more significant test cases, hybrid quantum annealing shows a relative error of less than 0.02\% compared to the classical benchmark and for IEEE 118-bus test case shows time performance speed-up. The increment in performance would enable real time planning and operations of electrical grids in real time. This work intends to be the first step to showcase the potentials of quantum computing towards the modernization and adaption of electrical grids to the decentralized future of energy systems.
\end{abstract}


\section{Introduction}
Across the globe, the share of electricity produced by Renewable Energy Sources (RES) is ramping up to address the challenge of power systems' decarbonization. Many consumers have transformed into prosumers in this context, generating their own electricity. 
However, the emergence of new actors poses a significant challenge to the Distribution System Operators (DSOs). RES have an intermittent and variable nature which increases the coordination efforts to ensure that the electrical grid remains safe and reliable in an ever more decentralized energy system \cite{RAPDSO}.

To accommodate more RES, DSOs need to upgrade the grid and improve the steering possibilities to provide enough flexibility to tackle these challenges. Partitioning the grid into logical clusters can help overcome some of these hurdles and leverage the new concepts of microgrids. It can incentivize the organization of local energy communities for energy exchange, prompting more consumers to engage as prosumers. 
Moreover, this clustering has proven beneficial in many technical aspects: improving energy balancing \cite{Koukaras2021}, more efficient use of reserves \cite{Chen6658958} and avoiding transaction leakage or loop flows, reducing the overall losses of the system \cite{4349093}. Also, logically creation of partitions can help analysing large power networks, avoiding an explosion of the computational and administrative costs \cite{Cotilla-Sanchez2013Multi-attribute}.

To optimally split the grid, many researchers have explored the applications of complex network and graph theory as a tool to reveal characteristics of the electrical grid and facilitate the clustering process, using modularity optimization as a key component.  

This approach has been applied from use cases in the transmission grid \cite{GUERRERO2018232,guerrero2019} and the distribution grid \cite{PAHWA20133741,Xu9119110,Xu2019Structural}. \cite{Xu9119110} and \cite{Xu2019Structural} have used graph theory approaches as a framework to adapt conventional distribution networks into virtual microgrids that could increase the flexibility of electrical systems. 
Other applications entail evaluating cascading failure risk, with the subsequent intentional islanding strategy \cite{PAHWA20133741}, or vulnerability assessment against other threats such as a terrorist attack \cite{WANG2017156}. This sort of analysis can help better understand the current risks of the grid to integrate new agents and systems of future decentralized grids.

However, obtaining grid partitions based on maximizing modularity is known to be a NP-complete combinatorial optimization problem \cite{NPComplete4358966}. In fact, in the references mentioned above, several methods were used from Hierarchical Clustering to Genetic Algorithms to overcome the complexity of the problem. In recent years, we have witnessed an interesting development in a new alternative to classical computation: quantum computing. Quantum computers use the properties of quantum mechanics to solve a particular set of problems such as combinatorial optimization potentially faster than conventional computers. Within the different architectures of quantum computing, D-Wave Systems (thereinafter D-Wave) developed the first commercially available quantum computers based on quantum annealing. 
The size of their Quantum Processor Units (QPU) and their hybrid methods library allows them to explore larger instances of problems \cite{D-WaveSystem0Welcome}. One can find examples of successfully using this technology in finance \cite{Rosenberg2016Solving} or logistics \cite{Ding2019Logistic}, among other areas. 
Moreover, in references \cite{Ushijima-Mwesigwa2017Graph,Negre2020Detecting}, researchers have successfully applied the concept of modularity-based graph partitioning based on quantum annealing. In their work, the authors were able to cluster and partition several networks by maximizing the modularity. However, not much work has been done in the intersection of quantum annealing and applications for the electrical grid.

This paper focuses on the application of quantum annealing to the partition of an electrical grid into logical communities, by maximizing the quality of the partition measured through modularity. The paper is organized as follows. Section \ref{sec:QA} will introduce quantum annealing. Section \ref{sec:gpqc} will cover the theoretical basis and the mathematical formulation of the binary optimization problem at hand. It introduces fundamental concepts of complex network theory and modularity, the basis of the solved binary optimization problem (graph partitioning through modularity optimization) explained below. 

Consequently, in Section \ref{sec:implement}, we will introduce the applied methodology and implementation procedure. This work uses both quantum annealing and hybrid methods from D-Wave Ocean's library \cite{oceansdk} to test the performance of quantum annealers. These methods are then benchmarked against two classical approaches: a well-known greedy algorithm for modularity optimization, the Louvain‘s algorithm \cite{JMLR:v21:20-412}; and an industry-standard integer programming solver, Gurobi Mixed Integer Programming (MIP) solver \cite{gurobi}. Section \ref{sec:results} will present three case studies we implemented based on the well-established IEEE benchmark. In Section \ref{sec:discussion} those results are discussed and lastly, Section \ref{sec:conclusions} provides the conclusions and outlines potential future work.

\section{Introduction to quantum computing}    
\label{sec:QA}

Quantum computing leverages quantum mechanical phenomena to process information. Within the different architectures, quantum annealing is the historically first applied use of quantum computing. Its sole application is to solve optimization problems by framing it as an energy minimization problem \cite{D-WaveSystem0What}. 

D-Wave quantum annealers implement superconducting qubits and they apply programmable external magnetic fields to influence their qubits (\textit{bias}) and \textit{couplers} to entangle those qubits. These biases and couplers are used to define an optimization problem. The energy state of this problem can be described mathematically through a Hamiltonian. The quantum annealing process is based on Schrödinger’s equation that determines the temporal evolution of a quantum state and the quantum adiabatic theorem \cite{Born1928Beweis}. 

The latter states that if a previously known ground state is picked and slowly varies to a final state (determined by a so-called target Hamiltonian), the resulting state will be the ground state of the target Hamiltonian. If the Hamiltonian is chosen carefully, it is possible to represent a particular type of optimization problem encoded in such a way that the ground/final state after quantum time evolution will represent the (possibly) optimum solution to the optimization problem \cite{Farhi2001quantum}. 

Therefore, the goal is to find a Hamiltonian function that represents a particular type of optimization problem. Thus, the Hamiltonian of the Ising spin glass model is selected \cite{Kadowaki1998Quantum}.

\begin{equation}
    \label{eqn:ising}
    H(s) = -1/2 \sum_{i,j} J_{ij} s_i^z s_j^z - H_z \sum_{i=1} s_i^z \;,
\end{equation}

where $s_i^z$ represents the magnetic dipole moments of the atomic spins, $J_{ij}$ the interaction coefficient between the spins and $H_z$ is the external magnetic field coefficient. The magnetic dipole can take the values of $1$ or $-1$. Each spin can be considered a qubit. Although a qubit is encoded by superposition of states, when measured, it collapses into one or the other state ($ | 0 \rangle$ or $ | 1 \rangle$). This is the reason why in optimization problems using quantum computers, the decision variables are binary.

Using the relationship $s_i^Z=2x_i-1$, an Ising problem can be expressed as a Quadratic Unconstrained Binary Optimization (QUBO) problem defined as: 

\begin{equation}
\label{eqn:QUBO}
\begin{gathered} 
    H(x)=x^T Qx	\;,
\end{gathered}
\end{equation}

where $Q_{ii}$  and $Q_{ij}$ are the diagonal and non-diagonal elements of Q matrix representing the problem and $x$ is a vector of binary decision variables $x_i$. 
The relationship between Equations (\ref{eqn:ising}) and (\ref{eqn:QUBO}) is relevant for two reasons. Firstly, it facilitates mapping it to combinatorial optimization problems. 

Secondly, combinatorial problems are computationally hard to solve and, in fact, many are NP-hard. Therefore, the possibility of mapping QUBO problems to the Ising spin glass model and applying the properties of quantum mechanics opens up a new opportunity. 

In fact, for modularity optimization graph partitioning, quantum annealing has shown very promising results \cite{Negre2020Detecting}. However, this method has not yet been applied to electrical grids. 

\section{Graph Partitioning using quantum annealing}
\label{sec:gpqc}
The following section explains the complex networks theory for community detection, describes the optimization problem and how to transform the Binary Integer Program (BIP) into QUBO one.

\subsection{Complex network theory and electrical modularity}
 
 One of the theory applied to optimal graph partitioning is the complex networks theory, which was introduced in \cite{Newman2004Finding} to explore community structures within large networks, such as the electrical grid. 
 
 A to-be-partitioned electrical grid can be represented as a graph defined as $G = (B,L)$ where set $B$ contains all the set of buses $i$ of the grid and set $L$ the electrical lines $(i,j)$. The number of buses and lines are described respectively as $n = |B|$ and $m = |L|$. Graph can contain more information than the topological properties. In the particular case for community detection in electrical grids, it is important to include information such as electrical distance, admittance or average power flow, as this information affect the behaviour of the system  \cite{Sanchez6774471}. 
In order to partition the graph representing the electrical grid into representative communities, we will maximize the quality metric of modularity. The modularity of the partition of a particular graph will indicate how strongly connected the elements of the partition are to respect of the other partitions. 

The concept of modularity is formally defined as \textit{“the number of edges falling within groups minus the expected number in an equivalent network with edges placed at random”} \cite{Newman2004Finding} expressed as follows: 

\begin{equation}
\label{eqn:modularity}
    Q = \frac{1}{2m} \sum_{ij} \left[ A_{ij} - \frac{k_i k_j}{2m} \right] \delta(C_i,C_j)\;,
\end{equation}

where $Q$ stands for the modularity, $m$ is the total number of edges in a graph, $A_{ij}$ is the coefficient for the $i,j$ th element of the adjacency matrix, $k_i$ the degree of bus  $i$ (the sum of the edges connected to $i$ ) and $\delta(C_i,C_j )$ is equal to $1$ if vertices $i$ and $j$ belong to the same partition and $0$ otherwise. 

Modularity has been used for several type of networks, including electrical grids. \cite{Xu2019Structural} introduced the concept of “electrical modularity” which is applied to assess the ideal partition of the grid at hand, while introducing weights related to the electrical properties of the grid:

\begin{equation}
\label{eqn:elecmodularity}
    Q_e = \frac{1}{2M} \sum_{ij} \left[ A_{ij}^E - \frac{A_{i}^E A_{j}^E}{2M} \right] \delta(C_i,C_j)\;,
\end{equation}

where $A_i^E$, and $M$ are defined as $A_i^E = \sum_{j \in B} A_{ij}^E$, and $M = 1/2 \sum_{ij} A_{ij}^E$. Each $A_{ij}^E$ is derived by the Electrical Coupling Strength (ECS) describing the electrical distance and the transmission capacity of the lines, using composite weights:
\begin{equation}
\label{eqn:adjECS}
    A_{ij}^E = \begin{cases}
        \overline{E_{ij}} = |\alpha\overline{Y_{ij}}+\beta\overline{C_{ij}}| & \forall (i,j) \in L\\
        0 & \text{otherwise}
       \end{cases}
\end{equation}
where $\overline{Y_{ij}}$ and $\overline{C_{ij}}$ are the weights that describe the electrical distances and the transmission capacities of the lines, respectively. $i$,$j$ are the buses belonging to the set $B$ of all buses. Both weights are normalized to scale them.
The average admittance $\overline{Y_{ij}}$ is calculated by obtaining the inverse of the electrical impedance matrix $Y_{ij}=1/Z_{ij}$ and normalizing it with the average $Y$. We have adopted the definition of electrical distance as in \cite{Cotilla-Sanchez2013Multi-attribute}.

The average coefficient of line sensitivity $\overline{C_{ij}}$ is calculated by finding the minimum value of the coefficient between the maximum power line transmission in a line $P_{(l max)}$ and the Power Transfer Distribution Factor (PTDF) on transmission line $l$ when there is an energy transaction between buses $i$ and $j$. $i$ and $j$ belong to the set $B$ of total buses and $l$ to the set $L$  of total lines. PTDF is a standard measure of line sensitivity used in power analysis calculations \cite{Guerrero2018Decentralized}. The composite weight coefficient allows us to compare the respective prioritization of one component over the other. This work uses the same value of $0,5$ to $\alpha$ and $\beta$.
After determining all the necessary weights to build the modularity matrix, the optimization problem is described. Let $x_{ik}$ be the binary variable of the problem, expressed as 1 if bus $i$ belongs to group $k$ and 0 otherwise; and $x_k = [x_{0k}, ..., x_{Bk}]$ the vector of the $B$ binary variables corresponding to the same $k$. The maximization of the modularity finds the optimal bus distribution in $k$ partitions, which present the highest modularity. A high modularity will imply a good community structure for the resulting partitions. The resulting optimization problem for $k$ partitions and generalizing Equation (\ref{eqn:elecmodularity}) is stated as follows \cite{Negre2020Detecting}:

\begin{equation}
\begin{aligned}
\label{eqn:maxqubo}
     \text{max} & &  \sum^K x_k^TQ_ex_k 
\end{aligned}
\end{equation}

\begin{equation}
\begin{aligned}
\label{eqn:constrainedprob}
    & \textrm{subject to} &  \sum^K x_{ik} =1 \quad\quad \forall i \in B
\end{aligned}
\end{equation}

This new formulation includes a new constraint to enforce that each bus is only assigned to one group (cf. Equation \ref{eqn:constrainedprob}). The number of binary variables $x_{ik}$ needed for the graph partition algorithm is equal to $(nK)$. It grows linearly with the number of nodes $n$ and the number of partitions $K$.

\subsection{From Binary Integer Programming (BIP) to QUBO}

In the previous section, the optimization problem has been defined. Mathematically, Equations (\ref{eqn:maxqubo}) and (\ref{eqn:constrainedprob}) represent a binary quadratic constrained problem. With the aim of solving this optimization problem using quantum annealing, this binary problem needs to be converted to a Quadratic Unconstrained Binary Optimization (QUBO) one.  This transformation is performed as follows. Let $H$ be the Hamiltonian given by

\begin{equation}
\label{eqn:hamiltonian}
    H=H_{obj}+\sum_i^B H_{c_i}\quad\text{where}\quad
\end{equation}

\begin{equation}
\label{eqn:hamiltonianobj}
   H_{obj}=-\sum^K x^TQ_e x\quad\text{and}\quad
\end{equation}

\begin{equation}
\label{eqn:hamiltonianc}
    H_{c_i}=\lambda  (\sum^K x_{ik}-1)^2\;,
\end{equation}
where $ H_{obj}$ is the objective function and
$H_{c_i}$ is the constraint Hamiltonian. As the original problem is a maximization, Equation (\ref{eqn:hamiltonianobj}) sign is switched in $ H_{obj}$. 
The only constraint is turned into a quadratic one, as seen in Equation (\ref{eqn:hamiltonianc}), where $\lambda$ is the penalty term. 

As QUBO is an unconstrained optimization problem, there are no “hard constraints” in comparison to Integer Programming (IP), Linear Programming (LP), or MILP. An enforcement's penalty term weights each constraint. Within the function, the penalty term quantifies the “importance” of the constraints. To the best of our knowledge, researchers and software providers agree that the choice of the terms is problem-dependent \cite{Glover2019Quantum} and often empirically set. 

\section{Implementation}
\label{sec:implement}
In this section, we will introduce the concept of quantum annealing and the methodology used to solve the optimization problem described above, in which we applied quantum annealing techniques.

As a preprocessing step, the $Q_e$ matrix must be calculated according to the procedure commented on Section \ref{sec:gpqc}. Afterwards, we adapted the formulation of problem described by Equations (\ref{eqn:maxqubo}) and (\ref{eqn:constrainedprob}) for the applied methods.  

To test the performance of D-Wave QPU, we used one quantum annealing sampler (\textit{DWaveSampler}) and two hybrid ones (\textit{LeapHybridDQMSampler}, and \textit{LeapHybridSampler}) from D-Wave's Ocean library \cite{oceansdk}. Whereas \textit{DWaveSampler} performs quantum annealing to solve the problem stated as the Hamiltonian (\ref{eqn:hamiltonian}), both \textit{LeapHybridDQMSampler}, and \textit{LeapHybridSampler} are hybrid methods that leverage both the capabilities of the QPU with state of the art D-Wave's proprietary classical algorithms. Hybrid approaches ease running problems with a higher number of variables and circumvents the limitations of the current state of the technology. 
Additionally, the hybrid \textit{LeapHybridDQMSampler} can efficiently accommodate the binary problem (cf. Equation \ref{eqn:maxqubo}) as a discrete one. Each variable $x_i$ representing a bus can take a range of values $K$. D-Wave's internal algorithms enforce that this condition holds, meaning that no node $i$ can be assign to more than one clusters. This conditions is exactly what Equation (\ref{eqn:constrainedprob}) describes. Consequently, one constraint and its $\lambda$ parameter tuning are avoided.  

\section{Case Study}
\label{sec:results}

This section illustrates the suggested partitioning method using three IEEE test case systems data. Both test cases are partitioned applying the methods described above, showcasing the results in terms of quality of solution and time performance.  

\subsection{Data Description}
This work tests the partitioning algorithm in two well-known IEEE test cases: IEEE 14-bus test case, IEEE 33-bus test case and IEEE 118-bus test case. Each test case contains a different number of power lines and buses as reported in Table \ref{table_example}. The IEEE 14 bus \cite{IEEE14_4075293} and the IEEE 188 bus \cite{case118} test cases exemplify a piece of the American Power System, and the IEEE 33 was introduced to represent a distribution network with switch control \cite{baran1989}. 
The used test cases data was obtained from the pandapower python package version 2.6.0 \cite{Thurner2018Pandapower}. 

\begin{table}[!h]
\renewcommand{\arraystretch}{1.0}

\caption{Number of elements in the IEEE 14-bus, IEEE 33-bus and IEEE 118-bus test cases}
\label{table_example}
\centering

\begin{tabular}{|c||c||c| |c|}
\hline
Element & IEEE 14 & IEEE 33 & IEEE 118 \\
\hline
Buses & 14 & 33 & 118\\
\hline
Lines & 15 & 37 & 173\\
\hline
Loads & 11 & 32 & 99\\
\hline
Generators & 4 & 0 & 53\\
\hline

Grid ext. & 1 & 1 & 1\\

\hline
Transformers & 6 & 0 & 13\\
\hline
\end{tabular}
\end{table}


\subsection{Results}

Table \ref{table_methods} summarizes the methods applied to each test case. 
All the results displayed in the following section have been conducted using an MacBook Pro with processor Intel® 8-Core™ i9 CPU @ 2.40GHz and 32 GB RAM. The code was run using Python version 3.8.8  and Jupyter Notebook 6.1.4. DWaveSampler has been run on D-Wave's Advantage 4.1. 5000 qubits QPU through its Leap cloud platform. Hybrid methods have been performed through the same platform with its corresponding hybrid solvers. 

\begin{table}[!h]
\renewcommand{\arraystretch}{1.0}

\caption{Summary of methods applied in each test case}
\label{table_methods}
\centering

\begin{tabular}{|l||c||c| |c|}
\hline
Method & IEEE 14 & IEEE 33 & IEEE 118 \\
\hline
DWaveSampler & \Checkmark & - & -\\
\hline
LeapHybridSampler & \Checkmark & \Checkmark & -\\
\hline
LeapHybridDQMSampler & \Checkmark & \Checkmark & \Checkmark\\
\hline
Louvain & \Checkmark & \Checkmark & \Checkmark\\
\hline

Gurobi & \Checkmark & \Checkmark & \Checkmark\\

\hline

\end{tabular}
\end{table}

\subsubsection{IEEE 14-bus test case}

Table \ref{modularity_results_table} depicts the maximum value of the objective function results obtained for different number of partitions $k$ and already mentioned methods. 

\begin{table}[!!!!h]
\caption{Objective function modularity for different partitions and different applied methods. }
\label{modularity_results_table}

\begin{tabular}{llllll}
                     & \multicolumn{3}{l}{Number   of communities k} &            &            \\[4pt]  \cline{2-6} \\[-0.5em]
Method               & 1            & 2             & 3            & 4          & 5          \\[3pt] \hline

\\ [-0.5 em]
DWaveSampler         & 0,000        & 0,3495        & \textbf{0,4646}       & 	\textbf{0,4844}     & 0,4393    \\

LeapHybrid    & 0,000        & 0,3495        & 0,4613       & 0,4613 & 0,4613  \\
LeapHybridDQM & 0,000        & 0,3495        & 0,4613       & 0,4613  & 0,4613  \\
Louvain              & -            & -             & 0,4613       & -          & -          \\
Gurobi (MIP)         & 0,000        & 0,3495        & 0,4613       & 0,4613  & 0,4613 
\end{tabular}
\end{table}

It is observed that all methods result into the same modularity value for the three partitions, where \textit{DWaveSampler} is able to obtain a valid partition with a higher modularity by 1\% (in bold in Table \ref{modularity_results_table})). After three partitions, \textit{LeapHybridSampler}, \textit{LeapHybridDQMSampler} and \textit{Gurobi's MIP Solver} reaches a maximum of the objective function value at $0.4613$. 
Conversely, row for \textit{DWaveSampler} shows that for exactly four and five partitions, the value for modularity decreases.

\begin{figure}[!!!!!!!h]
    \centering
    \includegraphics[width=0.4\textwidth]{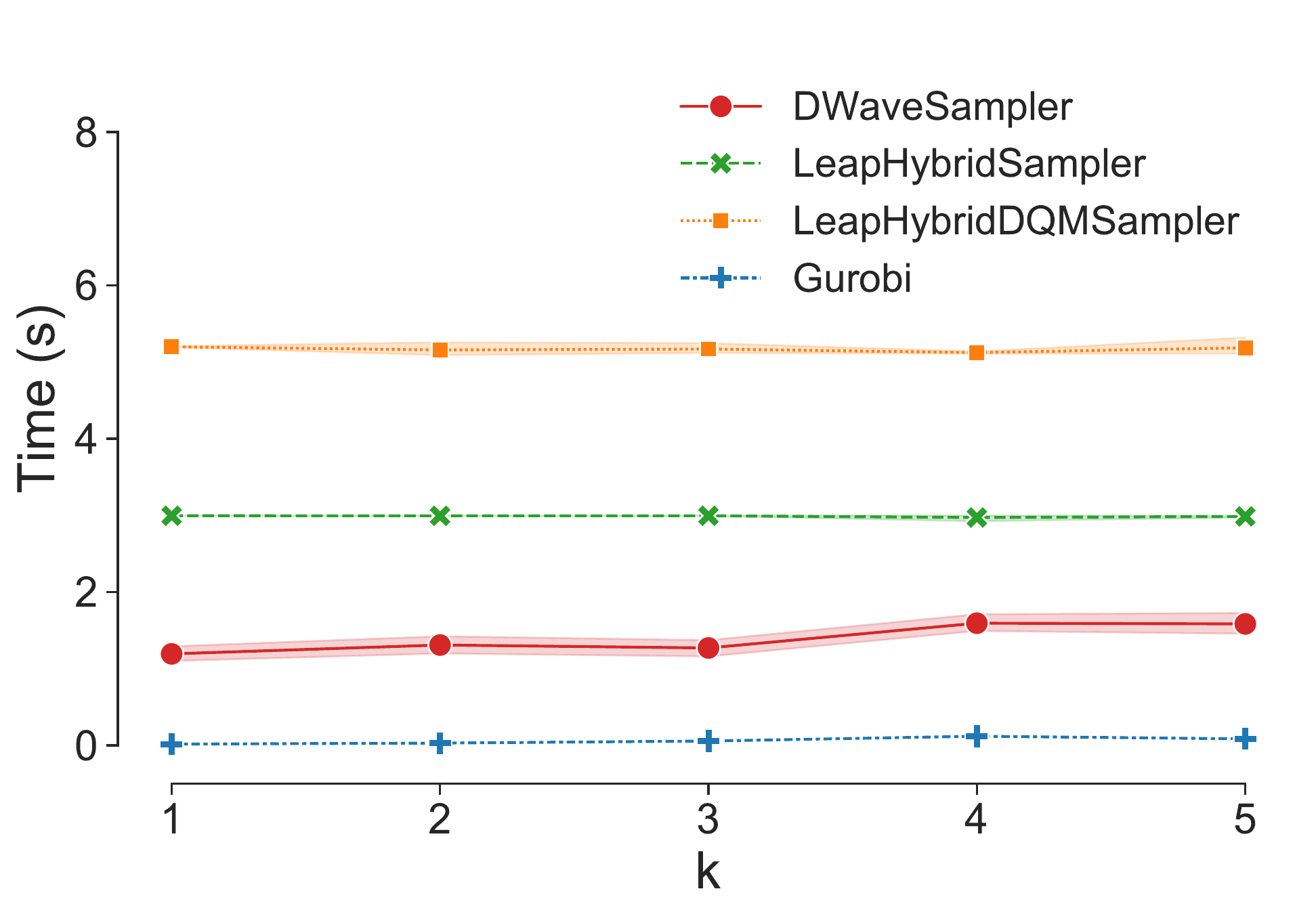}
    \caption{Average run time performance in seconds for each partition and tested method in the IEEE 14-bus test case}
    \label{fig:time14}
\end{figure}

Fig. \ref{fig:time14} depicts the average running times with their standard deviation for the tested methods. Gurobi MIP Solver shows to be faster than the quantum and hybrid methods by a difference between approximately 1 and 5 seconds, depending on the method. From the D-Wave's library, \textit{DWaveSampler} performs best with a running time under two seconds. 

Both hybrid samplers exhibit a constant running time. As stated by D-Wave's on resources, the minimum running time for \textit{LeapHybridSampler} and \textit{LeapHybridDQMSampler} are 3 and 5 seconds respectively, which can not be decreased. As the problem size is still under 1000 variables, a change in running time for these two methods is not discernible.

\subsubsection{IEEE 33-bus test case} In this case, we adopt the same procedure as in the previous subsection, not including \textit{DWaveSampler} in the analysis, as the problem size exceeds the qubit count of the current QPUs.

\begin{figure}[!h]
    \centering
    \includegraphics[width=0.4\textwidth]{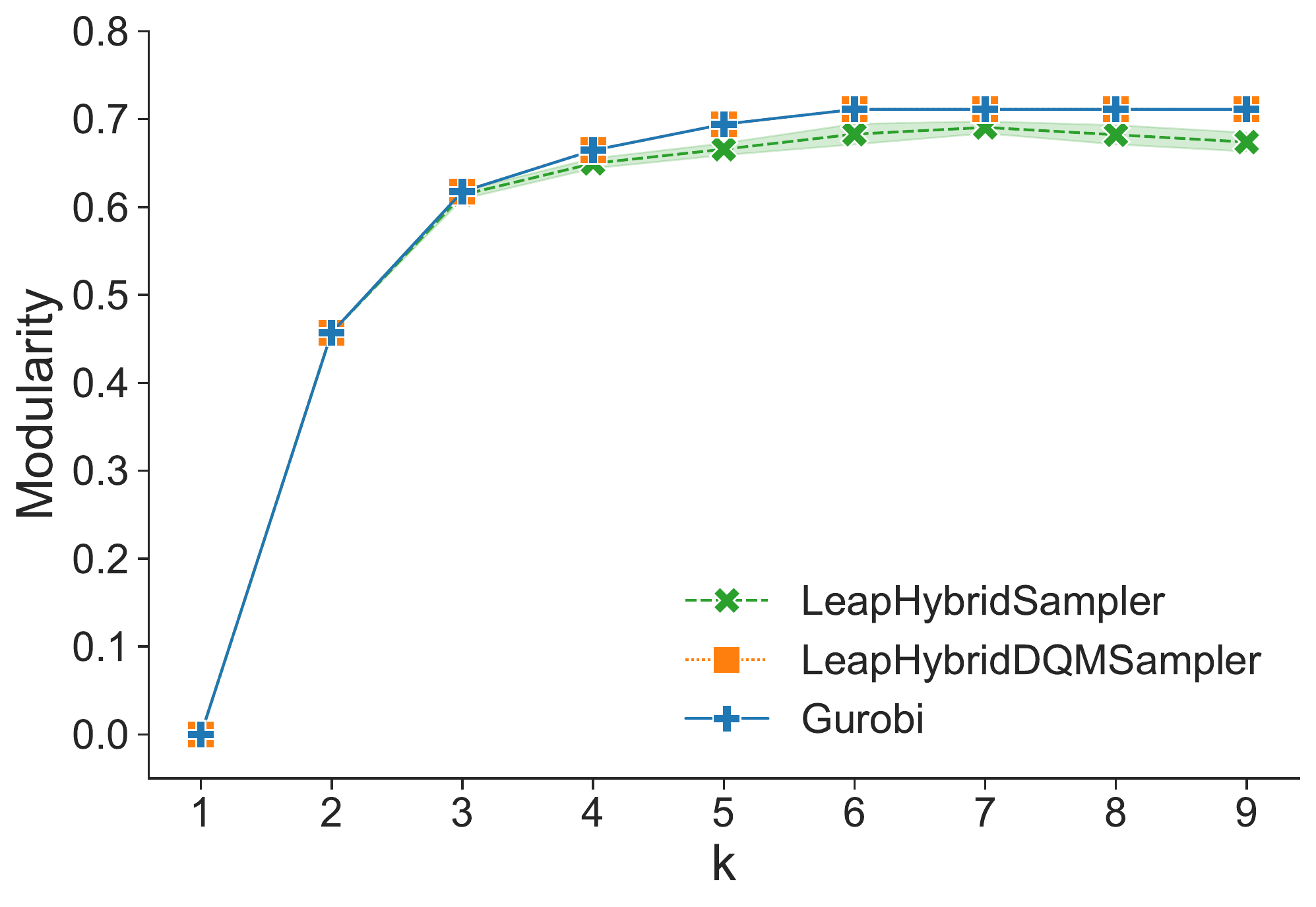}
    \caption{Modularity versus number of partitions plots for each tested method in the IEEE 33-bus test case.}
    \label{fig:mod33}
\end{figure}
\begin{figure}[!!!!!h]
    \centering
    \includegraphics[width=0.4\textwidth]{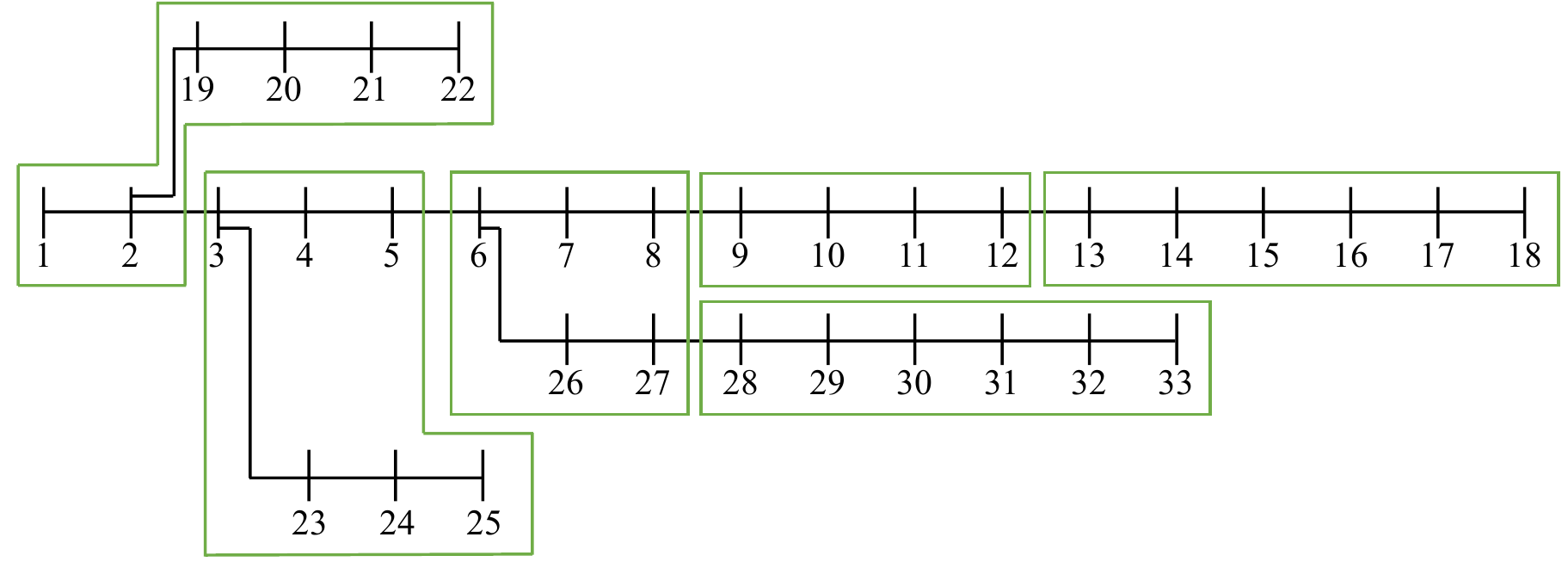}
    \caption{Partitioning results for the best modularity $(Q_e = 0,71)$ using LeapHybridDQMSampler and 6 partitions in the IEEE 33-bus test case.}
    \label{fig:partition33}
\end{figure}

Fig. \ref{fig:mod33} depicts the average results for different partitions of the IEEE 33 bus test case and the three applied methods. As observed, modularity reaches a maximum at $k=6$. This result is consistent with the one obtained with Louvain's algorithm with 6 partitions. Both \textit{Gurobi MIP Solver} and \textit{LeapHybridDQMSampler} achieve $Q_e = 0.711$, higher than the one obtained with greedy Louvain algorithm, $Q_e = 0.709$.
In Fig. \ref{fig:partition33} we can observe the obtained partition for the best modularity result applying \textit{LeapHybridDQMSampler}. 

\begin{figure}[h]
    \centering
    \includegraphics[width=0.4\textwidth]{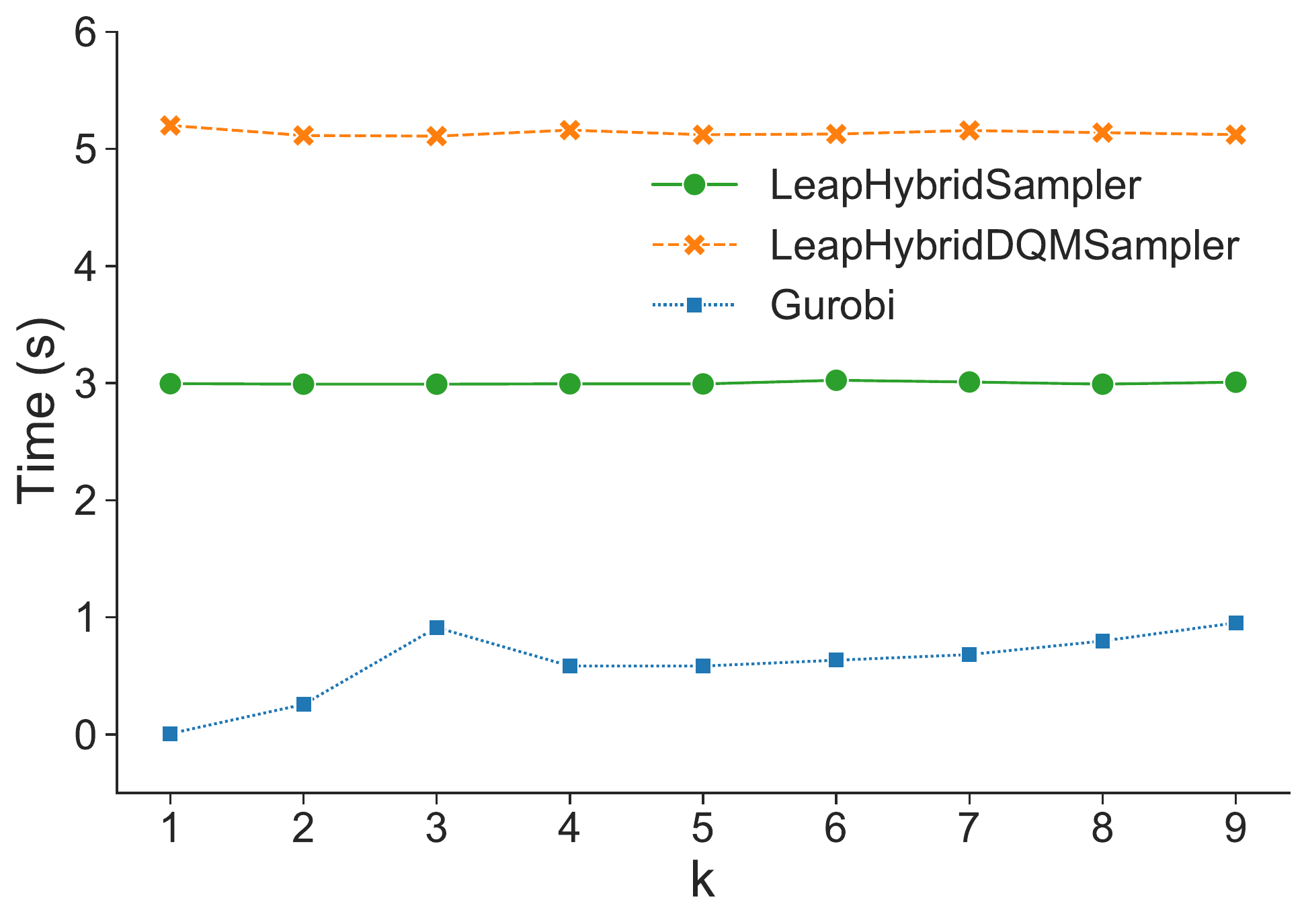}
    \caption{Average run time performance for each partition and tested method in IEEE 33-bus test case. }
    \label{fig:time33}
\end{figure}

When comparing the run time performance of each method, we observe a similar behaviour as in the test case used before. Gurobi MIP Solve runs the optimization problem in less time than the hybrid methods, as depicted in Figure \ref{fig:time33}. However, its running time behvaiour changes at $k=4$ (132 variables) and it increases with the number of $k$, as the number of variables is proportionate to $k$. As described in the previous section, both \textit{LeapHybridSampler} and \textit{LeapHybridDQMSampler} display a rather constant behaviour, following its time specifications. 

\subsubsection{IEEE 118-bus test case}
Table \ref{mv_results} depicts the results for 9 partitions in the IEEE 118-bus test case. Gurobi was not able to converge and time was stopped at the set time limit, 3600 s. However, result after convergence was higher than the LeapHybridDQMSampler but it is very close. 
\begin{table}[!!!h]
\renewcommand{\arraystretch}{1.0}
\caption{Summary of results of modularity and running time for selected methods and 16 partitions.}
\label{mv_results}
\centering
\begin{tabular}{|c||c||c| }
\hline
Method & Modularity & Running time (s) \\
\hline
LeapHybridDQMSampler & 0,7444 & 6,2 \\

\hline

Gurobi & 0,7448 & 3600 (*)\\

\hline

\end{tabular}
\end{table}

\begin{figure}[!h]
    \centering
    \includegraphics[width=0.4\textwidth]{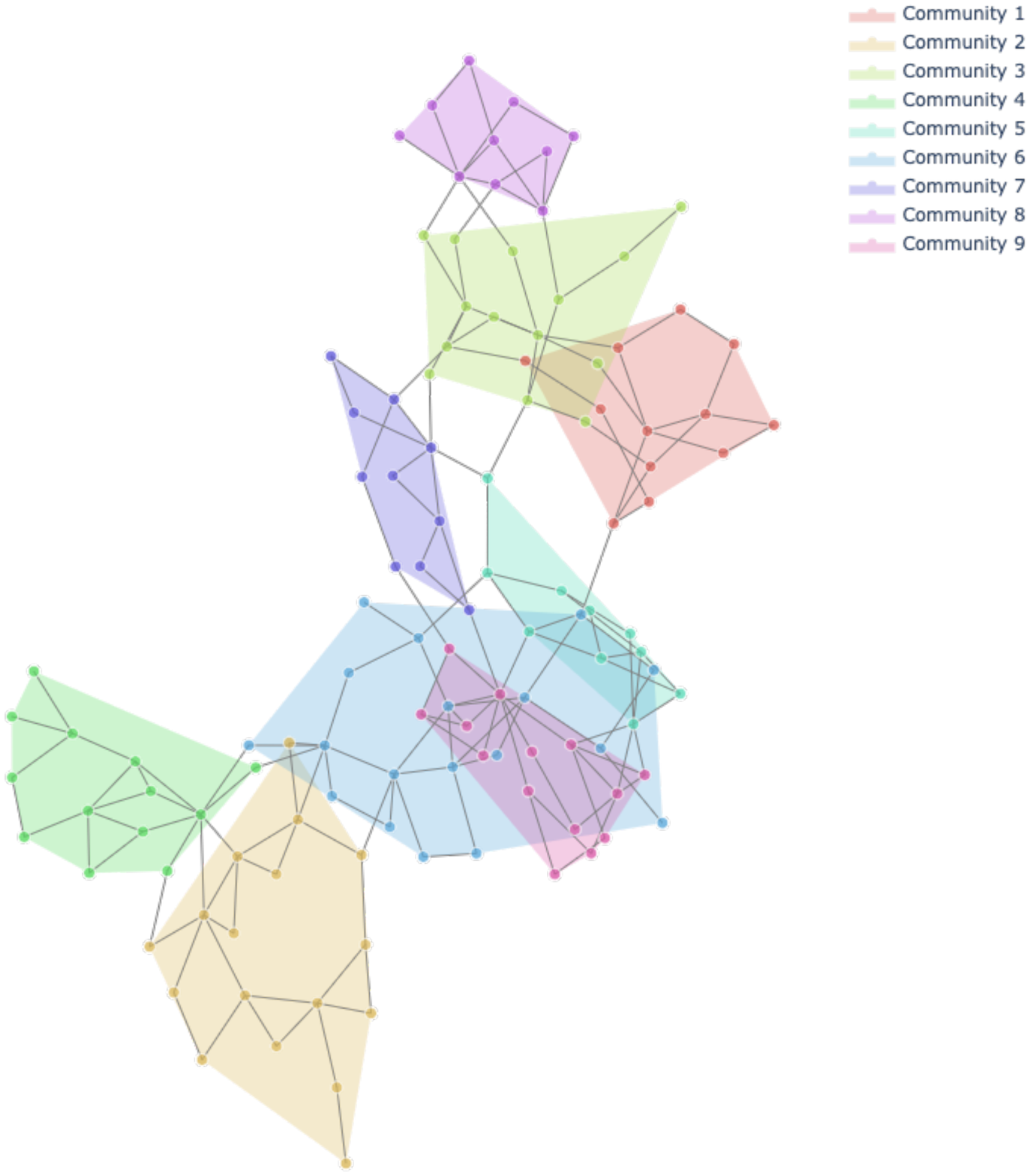}
    \caption{Partition of network IEEE 118-bus test case. }
    \label{fig:mv_partitiong}
\end{figure}

To visualize the results obtained with the \textit{LeapHybridDQMSampler}, Figure \ref{fig:mv_partitiong} depicts how the partition of the IEEE 118-bus test would results after performing the optimization algorithm. 

\section{Discussion}
\label{sec:discussion}

Results indicate that quantum annealing can show a potential speed-up for modularity optimization and it seems a suitable candidate to apply it for grid partitioning.

In the smaller test case, IEEE 14-bus, \textit{DWaveSampler} achieves a higher modularity result than hybrid counterparts or benchmarking classical method. For a medium grid, IEEE 33-bus, \textit{LeapHybridDQMSampler} achieves a performance comparable to the classical benchmark, Gurobi MIP Solver. 
The results might suggest that \textit{LeapHybridDQMSampler} is a superior method than \textit{LeapHybridSampler}. However, the difference for these results might come from the fact that whereas in \textit{LeapHybridSampler} the coefficient $\lambda$ for weighing equation \ref{eqn:hamiltonianc} is chosen by the user, \textit{LeapHybridDQMSampler} incorporates an optimization directly for this constraint. Since QUBOs are unconstrained models, there is always a trade-off between finding the optimal value for the objective function while ensuring the problem's conditions. Also, the current implementation of the D-Wave Discrete Quadratic models facilitates the implementation of such problems. These were the reasons why \textit{LeapHybridDQMSampler} was chosen for the larger network, IEEE 118-bus test case. 

In terms of time performance, classical methods exhibit a faster behavior than quantum and hybrid forms. However, \textit{DWaveSampler} samples the Hamiltonian 10.000 times to account for the probabilistic nature of quantum processors. This affects its performance as each annealing is set to last 20 $\mu s$. Therefore, the more samplings are conducted, the slower the running time is. For the hybrid methods, running time is limited by its specifications. Since the maximum number of variables needed for the IEEE test cases are relatively low (56 and 297 for IEEE 14-bus and IEEE 33-bus, respectively), the algorithms run at their minimum possible time: 3 seconds for the \textit{LeapHybridSampler} and 5 seconds for the \textit{LeapHybridDQMSampler}. 

It is not until a significant case with 1062 variables, IEEE 118-bus for nine partitions, that we observe a decrease in speed both in Gurobi MIP Solver and hybrid DQM sampler. Convergence with Gurobi was not achieved after an hour, which supports the claim that a problem such as a graph partitioning using modularity optimization quantum annealing could, potentially, show a quantum speed-up. Further research could compare these hybrid methods to other quadratic integer programming solvers.

\section{Conclusions}
\label{sec:conclusions}

This work provides a new framework to solve the graph partitioning problem through electrical modularity optimization using quantum annealing. It combines complex network theory with domain knowledge to create electrical coherent clusters by maximizing electrical modularity. 

We formulated the graph partitioning problem using modularity mathematically and transform it to QUBO. Several methods were applied for testing and benchmarking three test cases: IEEE 14-bus, IEEE 33-bus and IEEE 118-bus. The first one was able to be run applying directly quantum annealing. The larger data test cases were chosen in order to explore the scalability of the problem in terms of the number of variables. The size of this problem increases linearly with the number of partitions. 
Pure quantum as well hybrid annealing samplers produces results with a quality slightly better (+1\%) or equal (less than 5\%) than the classical benchmark, Gurobi MIP Solver. In terms of running time, both classical approaches outperformed quantum and hybrid samplers, except when the size of the problem increased. For the IEEE 118-bus test case, the exponential increase of combinations impeded Gurobi MIP from converging. 

This potential speedup would help to plan and operate electrical grids in real-time. Since the grid weights can be changed, one could explore the applications of self-supplied local energy communities to cluster consumers and prosumers based on energy needs and offers. With the increasing generation and emergence of prosumers, this type of analysis would need to be performed regularly. Additionally, the possibility of including static and dynamic information of electrical lines allows to bound the future decentralized energy communities to the characteristics of the electrical grids. The future research paths should explore the choice of different weights for the method and applications that could leverage the potential of this new technology.

Quantum computing and specifically quantum annealing is at its early stage of technology maturity. The current size and quality of the quantum chip, measured in the number of qubits and the performance of those qubits, restrict the type and size of problems that can be solved. However, with hybrid approaches like the one applied, researchers can start unlocking the potential of quantum annealing. 
Nonetheless, quantum annealing shows promising features that need to be investigated further. This work aims to be a first step into applying quantum computing research to the challenges of the future electrical grids. 

\section*{Acknowledgments}
The authors would like to thank Dr. Victoria Goliber for useful insights into tuning the D-Wave tools throughout the development of this project.

\bibliographystyle{IEEEtran}
\bibliography{IEEEabrv,bare_conf.bib}

\end{document}